\documentclass[notitlepage,nofootinbib,amssymb, nobibnotes, aps, prd,showpacs,showkeys]{revtex4-1}

\usepackage{graphicx}
\usepackage{hyperref}
\usepackage{epsfig}
\usepackage{fancyhdr}
\usepackage{empheq}
\usepackage{mathbbol}
\usepackage{wasysym}
\usepackage{color}
\usepackage{bbm}
\usepackage{bm}

\newcommand{{\xyz}}{$X$,~$Y$,~$Z$}

\begin{document}

\title{Production of  Tetraquarks at the LHC}
\author{A.L. Guerrieri$^\#$, F.~Piccinini$^\dag$, A.~Pilloni$^{*}$, A.D. Polosa$^{*}$}
\affiliation{\mbox{$^\#$Dipartimento di Fisica and INFN, Universit\`a di Roma `Tor Vergata', 
Via della Ricerca Scientifica 1, I-00133 Roma, Italy}\\
$^\dag$INFN Pavia, Via A. Bassi 6, I-27100 Pavia, Italy\\
\mbox{$^*$Dipartimento di Fisica and INFN, `Sapienza' Universit\`a di Roma, P.le Aldo Moro 5, 
I-00185 Roma, Italy}
}

\begin{abstract}
\noindent
\newline
The  high prompt  production cross section of $X(3872)$ at hadron colliders 
has shown to be very informative about the quark nature of the {\xyz} states. 
We present here a number of results on $X$ production in $pp(\bar p)$ collisions 
obtained  with Monte Carlo hadronization methods and illustrate  what can be learned 
from their use to improve our understanding of exotic states.  In particular, a comparison 
between antideuteron and $X$ production cross sections is proposed. Hadronization  
might be the key to solve the problem of the extra states  expected in diquark-antidiquark 
models which are naturally favored after the recent confirmation  of the $Z(4430)$ tetraquark, 
together with its lower partners $Z_c(3900)$ and $Z_c^\prime(4020)$.
\end{abstract}
\maketitle

\section{Introduction}
The $X(3872)$ has been observed with very large ($\approx 30$~nb) production cross sections 
both at CDF~\cite{cdfprompt} and CMS~\cite{cmsprompt}. This result is  at odds with a 
loosely bound, $E_b\approx 0$, molecular description  resulting from the faint 
attraction of a $D^0$ by a $\bar D^{*0}$ meson~\footnote{The charge conjugated modes are understood.}. 
Such a na\"ive  statement was confirmed by numerical simulations in~\cite{bigna1,bigna2} 
and~\cite{arto}: an upper bound for the prompt production cross section at CDF was estimated 
to be $300$ times smaller than the experimental value. 

However in~\cite{arto} it was proposed that final state interactions {\it within} 
the $D^0\bar D^{*0}$ pair could make the pair coalesce into a hadron molecule even if the 
recoil momentum  in the center-of-mass (com) of  the two components were as large as 
$k_0\sim 3 m_\pi\simeq400$~MeV, compared to  the value $k_0\approx 50$~MeV compatible with the 
experimental binding energy $E_b = (-0.142 \pm 0.220)$~MeV~\cite{kamseth}. 
To make a comparison, consider that the deuteron, which is often addressed as the baryon 
analog of the $X(3872)$,  has $k_0\approx 80$~MeV. More studies followed in the field along 
the lines suggested in~\cite{arto}, see for example~\cite{altri} for a most recent account.

In~\cite{pioni} we reconsidered the problem of treating final state interactions as  was done in~\cite{arto} and, following a line of reasoning initiated in~\cite{bigna2}, we suggested to use  hadronization pions comoving with $D^0\bar D^{*0}$  pairs to elastically rescatter one $D$ or $D^*$ in such a way as to decrease the com  $k_0$ thus feeding  the number of would-be loosely bound molecules. The very existence of such comoving pions, that  we naturally have in numerical simulations, makes questionable~\cite{bigna2} the way  to resort to final state interactions within the $D^0\bar D^{*0}$ system adopted in~\cite{arto,altri}. 
 
However not even the method proposed in~\cite{pioni} is found  to be effective at explaining the very large prompt production cross sections observed for the $X(3872)$ at hadron colliders.

Very  recent experimental results indicate clearly the existence of genuine tetraquark mesons as the charged  $Z(4430)$, just confirmed by the LHCb collaboration~\cite{z4430}, and the $Z_c(3900)$  discovered by BES and Belle~\cite{bes3}. There are strong reasons to believe that these resonances could be all of a kind, and the challenge for theory is that of finding descriptions explaining the general pattern which emerges from available data. 

The qualitative prediction of such charged states was provided by a tetraquark model years ago~\cite{xmai,tetraz,xmai2}.  An updated version of that model has been presented in~\cite{wiprog}, where a unified description of the $Y$'s resonances with quantum numbers $J^{PC}=1^{--}$ (including the charmed baryonium discussed in~\cite{coto}), together with the $Z_c$'s  and $Z(4430)$ ($1^{+-}$ resonances), and the $X$ ($1^{++}$), is proposed. We claim here that  the `extra', not observed states, also predicted by the `type-II' tetraquark model, might  be  forbidden at the hadronization level. Here we focus on prompt production in hadron collisions and not on production in $B$ decays. 

{\bf\emph{Summary of results}}
$i)$ We show that elastic scatterings with comoving pions {\it do not} distort high transverse momentum cross section distributions: therefore they  provide  a viable way to study the potential reduction of com $k_0$ in $D$ meson pairs through final state interactions.
$ii)$ We discuss our simulations using $c\bar c$ production in HERWIG~\cite{her}, the $c\bar{c}g$ production in ALPGEN~\cite{alp} with a hard recoiling gluon, interfaced with HERWIG, and a full QCD production using HERWIG, where all possible partonic subprocesses with light quarks and gluon contributions are considered. The differences found are instructive about the use of these techniques for the study of molecule production at hadron colliders.
$iii)$ We find that final state elastic rescatterings with one pion at a time are  not sufficient to enhance the $X$ prompt production cross sections up to the observed values. On the other hand, since we find that we cannot exclude rescattering with comoving pions, a consequence is that the 
cross sections estimated with the method proposed in~\cite{arto} should be modified. This information can be of use to studies like those in~\cite{altri}.
$iv)$ No more than two or three pions {\it per event} are shown to be potentially effective at  coalescing the molecule. The overall effect of final state interactions with comoving pions gets larger when a sequence of $2\div3$ successive $\pi D$ scatterings is allowed,  making the considerations in the previous point even more evident. Anyway,  considering a stochastic sequence of elastic scatterings with pions also does not solve the prompt production problem if we integrate in the natural $k_0\in[0,50]$~MeV range.
$v)$ We discuss for the first time the production of antideuteron at LHC using some preliminary data analysis provided from the ALICE collaboration. We show how the remarkable production of antideuteron at very low transverse momentum is Monte Carlo (MC) extrapolated up to those $p_T$ values where the $X(3872)$ is observed to be copiously produced at CMS~\cite{cmsprompt}.
$vi)$ Recent experimental findings corroborate the bases of the compact tetraquark interpretation of {\xyz} states. We suggest that hadronization might have a role in 
forbidding those `extra' states also predicted by the type-II tetraquark model in~\cite{wiprog}.

\section{Tuning Hadronization Monte Carlo on CDF Data}
CDF  measured the $d\sigma/d(\Delta\phi)$ distribution for pairs of $D^0D^{*-}$ mesons produced  with a relative $\Delta\phi$ angle in $p\bar p$ collisions at Tevatron~\cite{doubleD}. We use this distribution to tune the normalization of the HERWIG hadronization algorithm. The same normalization is used to generate $D^0\bar{D}^{*0}$ pairs.  

\begin{figure}[h]
\centering
\includegraphics[width=.48\textwidth]{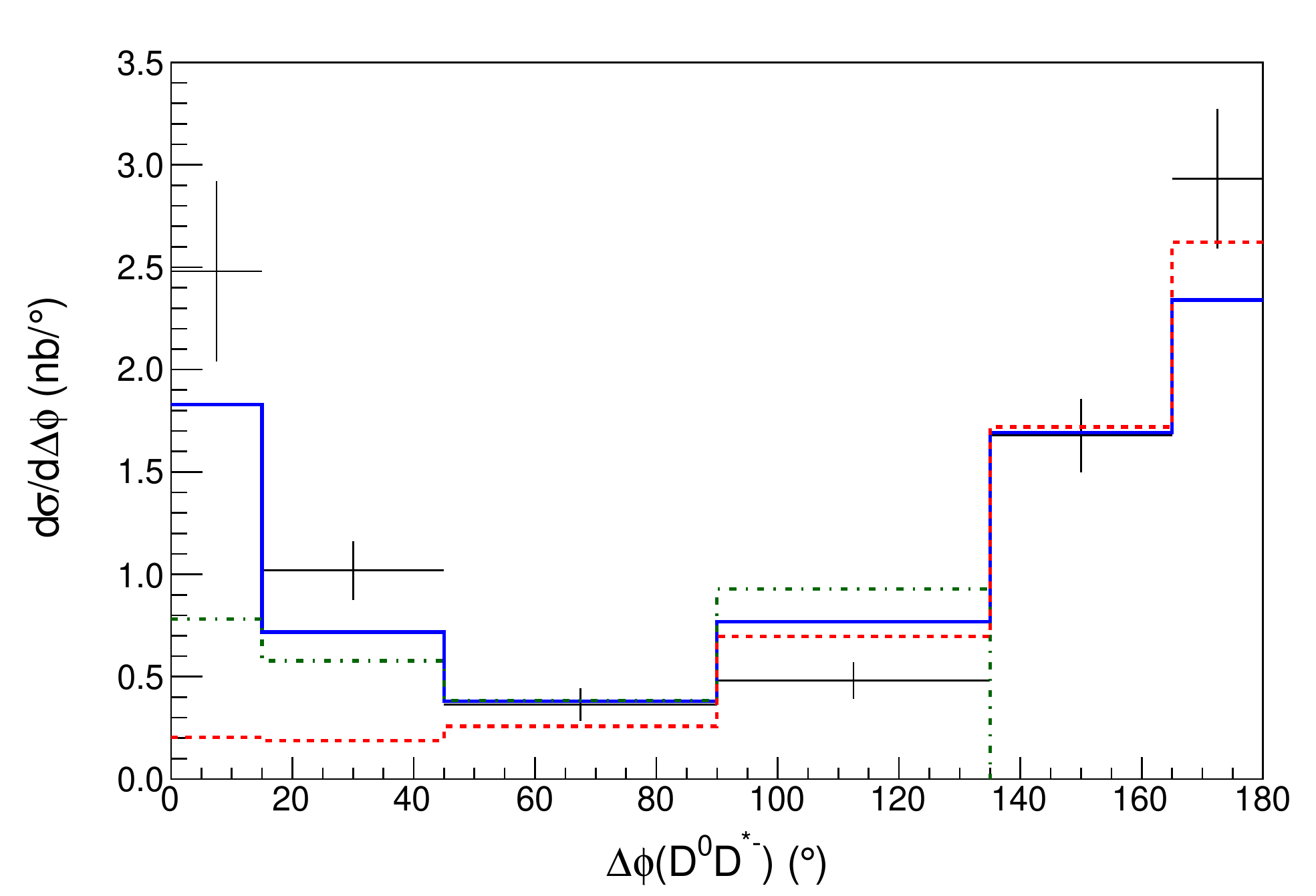}
 \caption{Differential cross sections of $D^0D^{*-}$ at CDF, with $2\to 2$ full QCD (blue, solid), with $2 \to c\bar c$ (red, dashed) and with $gg\to c\bar c g$ (green, dot-dashed).}
 \label{prima}
\end{figure}

With reference to Fig.~\ref{prima}, we  distinguish between different production methods: $i)$~Generate  $6\times10^9$, $2\to 2$ full QCD events  (blue, solid), and $ii)$~ $6\times10^8$ events of the type $2 \to c\bar c$   (red, dashed)  by parton shower, with $p_T^\text{part} > 2\text{ GeV}$, $\left|y^\text{part}\right| < 6$. Moreover, we generate $iii)$~$4\times10^8$ events of the $gg\to c\bar c g$ type  with a matrix element calculation (ALPGEN) requiring $p_T^{c,\bar c} > 3.5\text{ GeV}$, $p_T^{g} > 2\text{ GeV}$, $\left|y^\text{part}\right| < 2$. In order to fit data, we rescale the $2\to2$ MC distributions by a $K$-factor which minimizes the $\chi^2$ function. 
The distribution obtained with $gg \to c \bar c g$ (green, dot-dashed) is not reliable at large values of $\Delta\phi$, so we decide to show only the first four bins, normalizing this curve according to the central bin $[45^\circ,90^\circ]$.
The technique used here is that introduced 
in~\cite{bigna1}. 

We see that the complete QCD simulation well describes the whole $\Delta \phi$ experimental  distribution, whereas the other ones fail in the higher/lower parts:
consider that the low $\Delta\phi$ bin is the one with the most would-be-molecule candidates. 
Comparing the blue solid line histogram 
with the green dot-dashed one, we can infer that a large contribution 
comes from events which do not contain a $c \bar c$ pair at the matrix 
element level. Therefore the contribution of 
$g \to c \bar c$ conversions in the shower is relevant and can only be taken into account with 
a full QCD simulation. Even if this contribution is very sensitive 
to the hadronization model, we checked that PYTHIA~\cite{pythia} and HERWIG give 
similar predictions. 
Here we will limit our discussion to full QCD simulations. 

\begin{table}[h]
\begin{tabular}{c|c|c}
  & $K$-factor & $\chi^2/\text{DOF}$ \\ \hline
 full QCD & $1.65$ & $19.9/5$\\
 $c\bar c$ & $3.53$ & $68/5$\\
 $c\bar c g$ & $5$ & $-$\\ \hline
\end{tabular}
\caption{The MC normalization $K$-factors and $\chi^2$ relative to Fig.~\ref{prima}. The $2\to c\bar c g$ are normalized by constraining the value of the central bin.}
\end{table}

{\bf \emph{Interaction with pions at CDF}}
Given the large number of pions produced in the momentum phase-space neighbourhood of the open charm meson pairs (we call them `comoving' pions), it is plausible that some of  those  could scatter  elastically on the $D^0$ or $D^{*0}$ component of the would-be-molecule  changing the relative momentum in the centre of mass of the pair, $|\bm k_0|$,  towards lower values. We can assume the initial total energy $E$ of the pair to be positive. However, if  $k_0=|\bm k_0|$ gets smaller due to an interaction with the pion, $E$ might be found shifted downwards to some {\it negative} -- close to zero -- value, provided that the $D^0\bar D^{*0}$ pair is under the influence of an attractive potential, say a square well
potential, similar to the simplest description of deuterium.

In this way the $X$ would then be a genuine, negative energy, bound state of $D^0\bar D^{*0}$ whose lifetime is entirely regulated by the lifetime of the shorter lived component $D^{*0}$. In that case we would estimate then a total width  \mbox{$\Gamma_{\mathrm{tot}}(X)\simeq 65$~keV}. There are no energetic arguments to stabilize the $D^*$ in the attractive potential.

The elastic interactions with the pions are regulated in the $\pi D$ centre of mass  by the matrix elements
\begin{eqnarray}
&& \langle \pi(p) D(q) | D^*(P,\eta)\rangle = g_{_{\pi D D^*}}\, \eta \cdot p\\
 &&\langle \pi(p) D^*(q,\lambda) | D^*(P,\eta)\rangle = \frac{g_{_{ \pi D^* D^*}}}{M_{D^*}}\, \epsilon_{\alpha\beta\gamma\delta} \lambda^{\alpha} \eta^{\beta} p^{\gamma} q^{\delta}
\end{eqnarray}
where the couplings used are $g_{_{\pi D D^*}} \approx 11$, $g_{_{\pi D^* D^*}} \approx 17$, see~\cite{casal}.
After the interaction with the pion has taken place in the center of mass $D^0\pi$ frame, we boost back the $D^0$ in the laboratory (lab) frame and check if the `new' $D^0\bar D^{*0}$ pair passes the cuts we fixed for the final meson pairs. 

In this sense the mechanism proposed here is different from the one based on the assumption that the $D^0\bar D^{*0}$ pair should rescatter remaining  isolated from other comoving hadrons~\cite{arto,bigna2}. 

We start from the same setup discussed in~\cite{pioni,bigna1,bigna2}: $p \bar p$ collisions, $\sqrt{s}=1.96$~TeV. Each $D^0$ within loose cuts $1\text{ GeV} < p_T < 30\text{ GeV}$, $\left|y\right|< 4.5$ interacts with a pion; then we apply the experimental cuts $5.5\text{ GeV} < p_T < 20\text{ GeV}$, $\left|y\right|< 1$. 
The interacting pions are chosen randomly among those having $|\bm k_{\pi}|<M_D^{0(*)}$, so that we ensure they are close in phase space.
We normalize the MC by minimizing the $\chi^2$ to both the single-$D$~\cite{singleD} and double-$D$~\cite{doubleD} distributions provided by CDF. Finally, we normalize the MC to the $\Delta\phi$ distribution. 

\begin{figure}[htb!]
\centering
 \includegraphics[width=.48\textwidth]{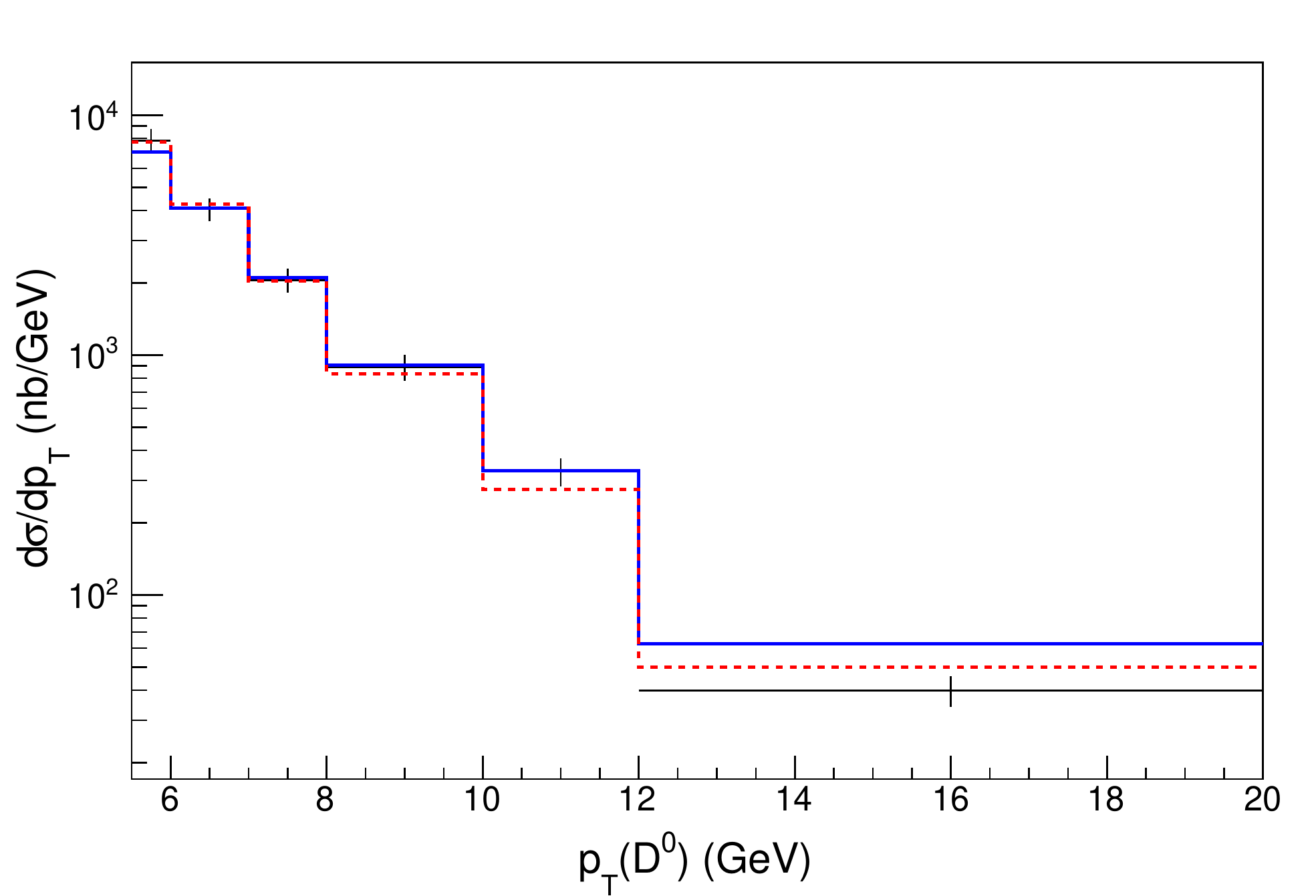}
 \includegraphics[width=.48\textwidth]{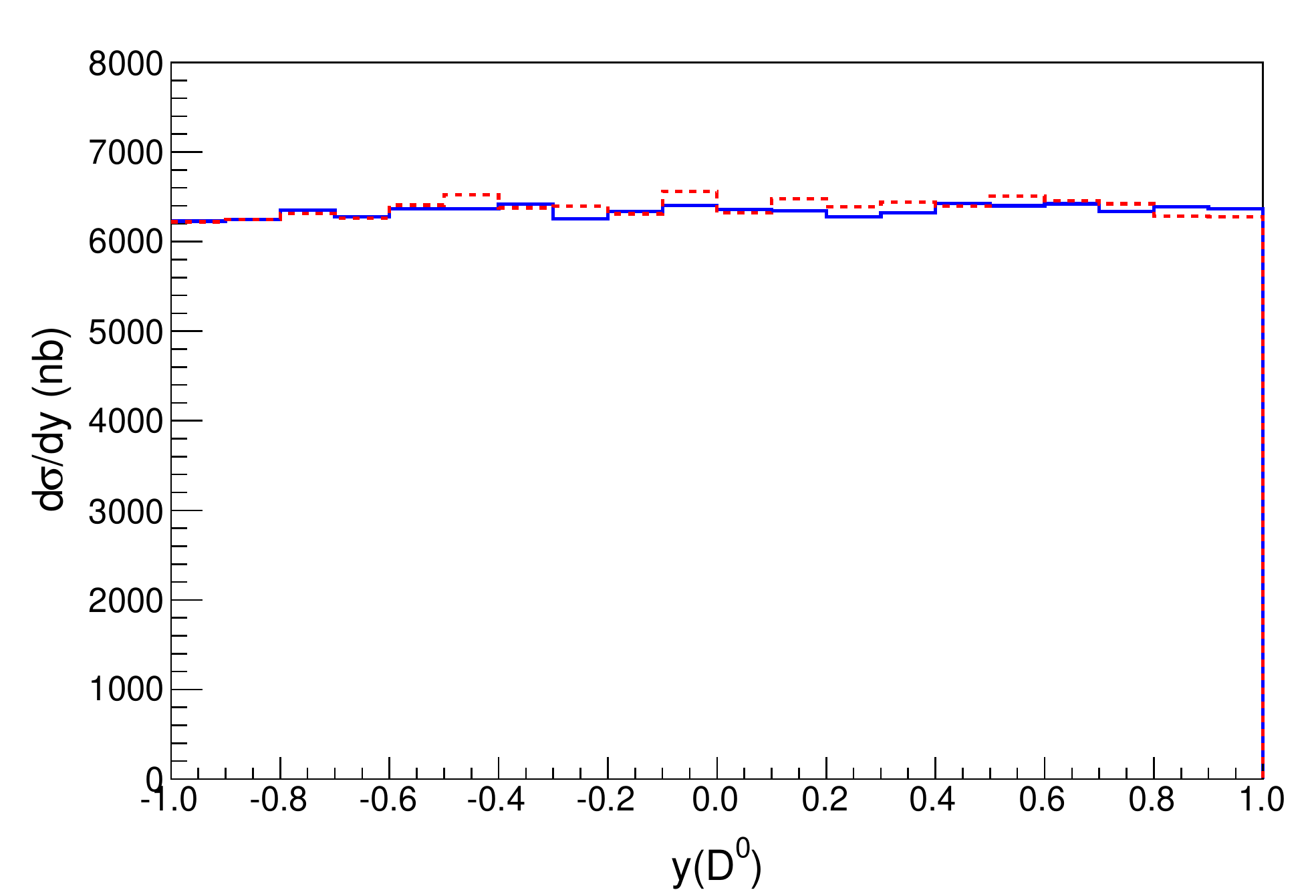}\\
 \includegraphics[width=.48\textwidth]{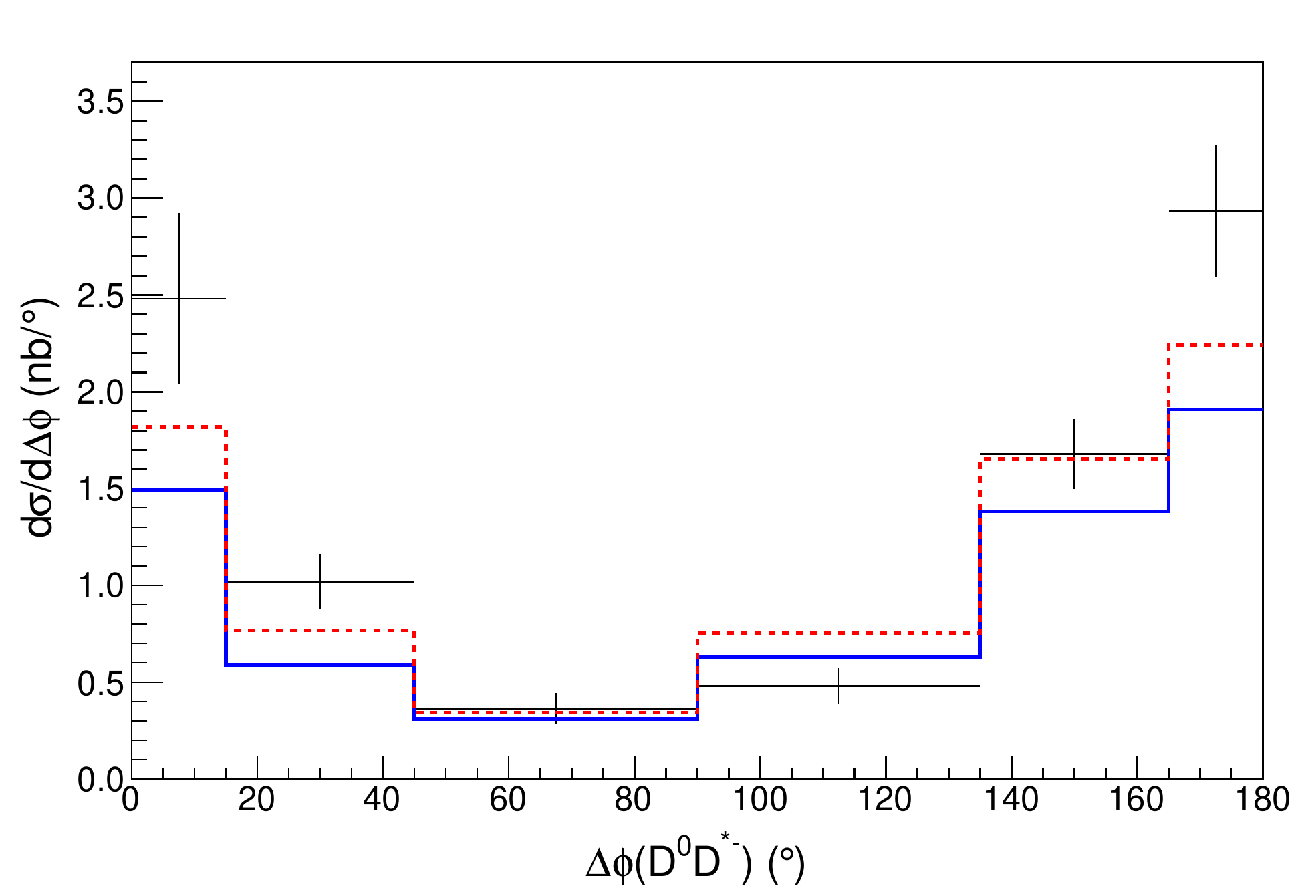}
 \includegraphics[width=.48\textwidth]{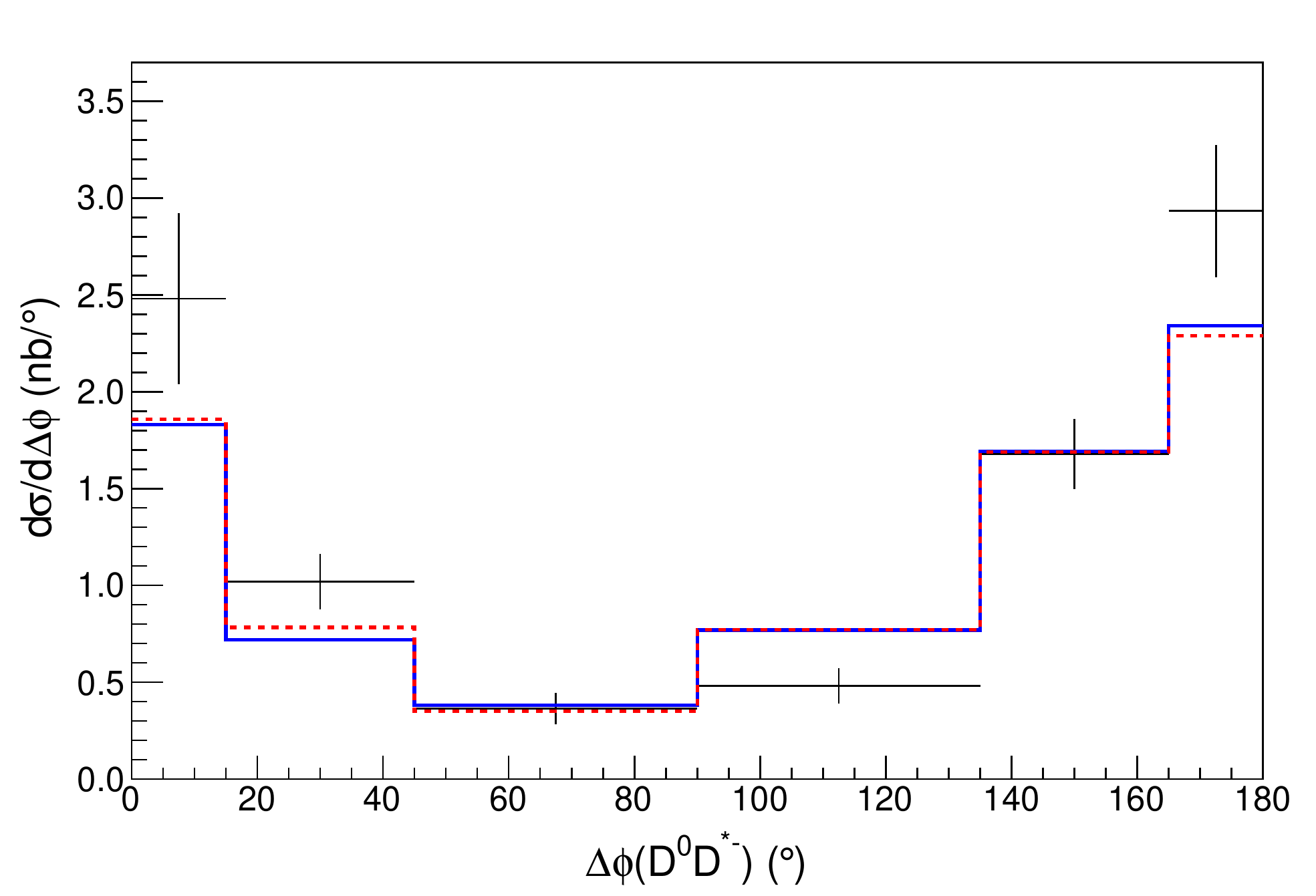}
 \caption{Differential cross sections of $D^0$ and $D^0 D^{*-}$ pairs at CDF, not including (blue, solid) and including (red, dashed) the interaction with one pion per $D^0$/$D^0D^{*-}$ event. The single-$D$ distributions and the left $\Delta\phi$ distribution have the same $K$-factor, obtained by minimizing the combined $\chi^2$. In the lower right panel we rescale the distribution in order to fit only the $\Delta\phi$ data. What is evident from this distributions is that the inclusion of an elastic scattering with a final state comoving pion does not spoil the high energy behavior in the transverse momentum and does not weaken the agreement with $d\sigma/d\Delta\phi$ distribution of $D^0D^{*-}$ pairs at CDF. No data are available on rapidity, yet pion scattering in the final state does not harm the distribution with respect to what found not including it. }
 \label{normaz}
\end{figure}

As it is illustrated in Fig.~\ref{normaz}, the inclusion of the interaction with a pion in the final state, both with a $D^0$ ($D^{*0}$) meson or with a $D$ component of the would-be hadron molecule, {\it does not spoil} the high energy behavior of the relevant distributions.  Fit values referring to Fig.~\ref{normaz} are reported in Table~\ref{table1}.

\begin{table}[htb!]
\begin{tabular}{c||c|c||c|c|}
 & \multicolumn{2}{|c||}{Combined fit} & \multicolumn{2}{|c|}{$\Delta\phi$ only} \\ \hline
  & $K$-factor & $\chi^2/\text{DOF}$ & $K$-factor & $\chi^2/\text{DOF}$\\ \hline
 $0\pi$ (blue) & $1.35$ & $45/11$ & $1.65$ & $19.9/5$\\
 $1\pi$ (red) & $3.46$ & $24/11$ & $3.53$ & $18.6/5$\\ \hline
\end{tabular}
\caption{Fit values referring to Fig.~\ref{normaz}.}
\label{table1}
\end{table}

\begin{figure}[h]
\centering
 \includegraphics[width=.48\textwidth]{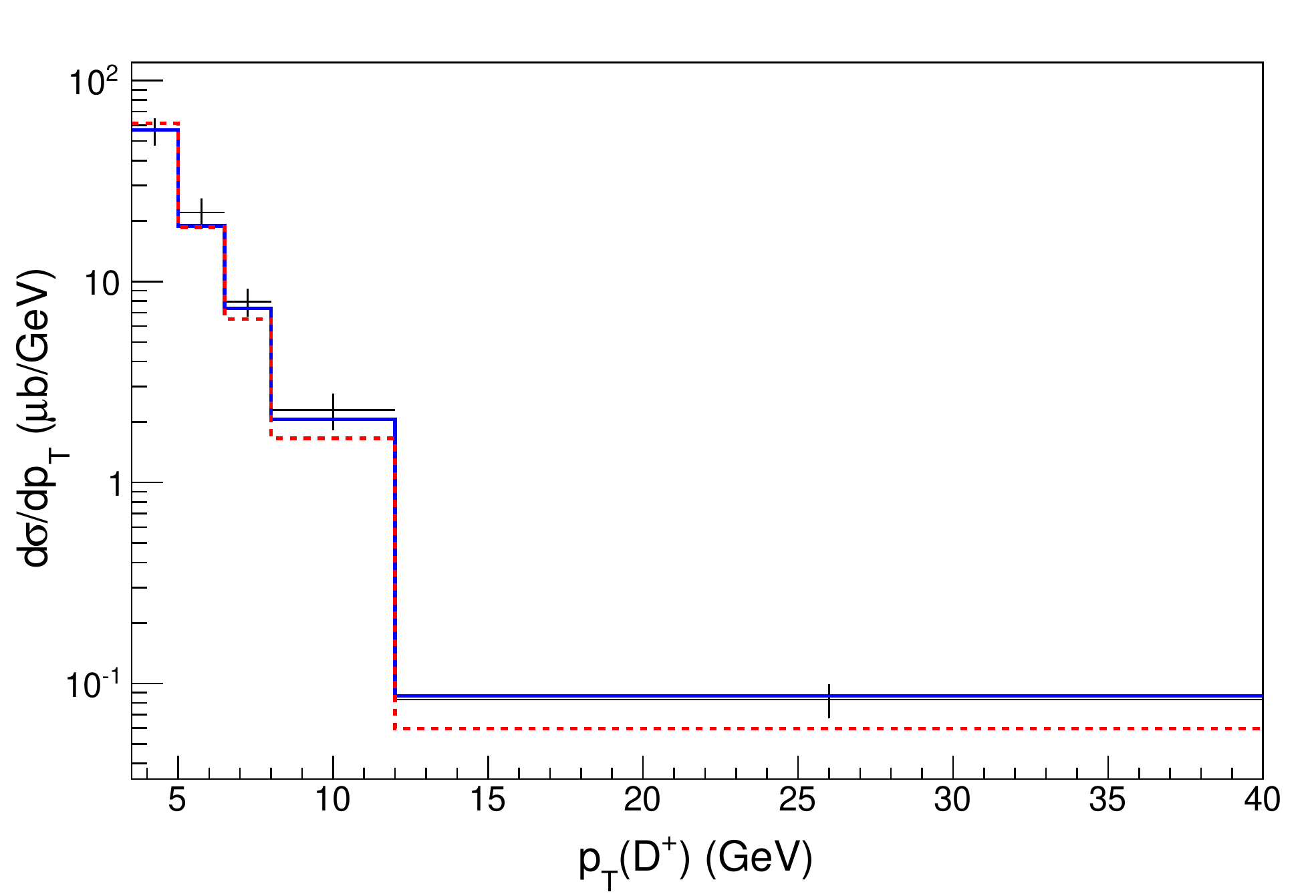}
 \includegraphics[width=.48\textwidth]{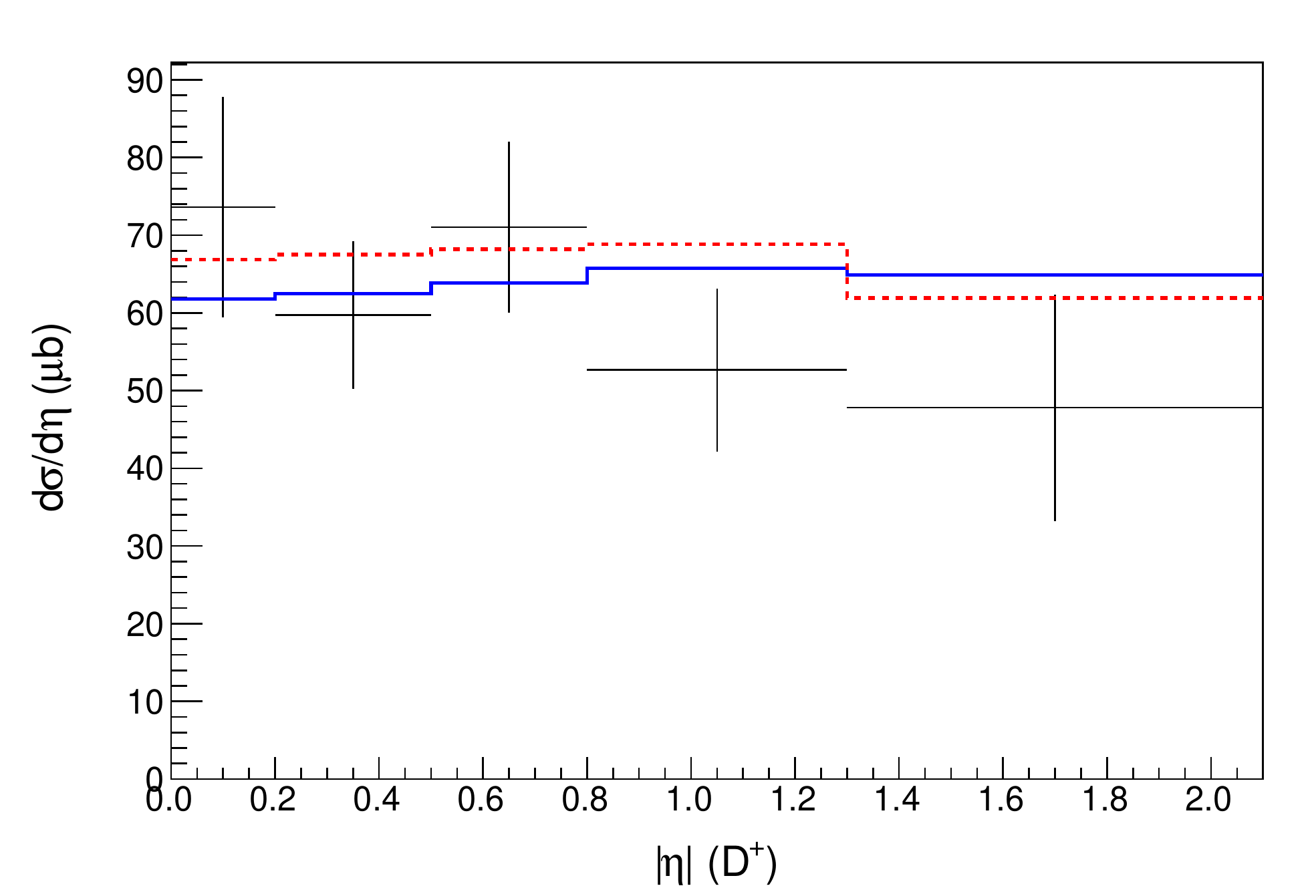}\\
  \caption{Differential cross sections of $D^+$ at ATLAS, not including (blue) and including (red) the interaction with final state pions. }
\label{figatlas}
\end{figure}

{\bf \emph{Interaction with pions at ATLAS}}
We use ATLAS data~\cite{ATLAS} on differential cross sections of $D^+$ mesons to check whether the potential deformations induced by the interaction with pions increase at higher energies. We let each $D^+$, within loose cuts $1\text{ GeV} < p_T < 50\text{ GeV}$, $\left|\eta\right|< 6$, interact with a pion. Then we apply the experimental cuts $3.5\text{ GeV} < p_T < 40\text{ GeV}$, $\left|\eta\right|< 2.1$. We normalize the MC to the single-$D$ ATLAS distribution~\cite{ATLAS}. We report in Table~\ref{tabatlas} the numerical results found for the fit in Fig.~\ref{figatlas}. We do not find appreciable differences with the discussion related to Tevatron energies.

\begin{table}[h]
\begin{tabular}{c|c|c}
  & $K$-factor & $\chi^2/\text{DOF}$ \\ \hline
 $0\pi$ (blue) & $0.45$ & $5.3/9$\\
 $1\pi$ (red) & $1.01$ & $10.6/9$\\ \hline
\end{tabular}
\caption{Fit values referring to Fig.~\ref{figatlas}.}
\label{tabatlas}
\end{table}

\section{Discussion}
\begin{figure}[htb!]
\centering
 \includegraphics[width=.48\textwidth]{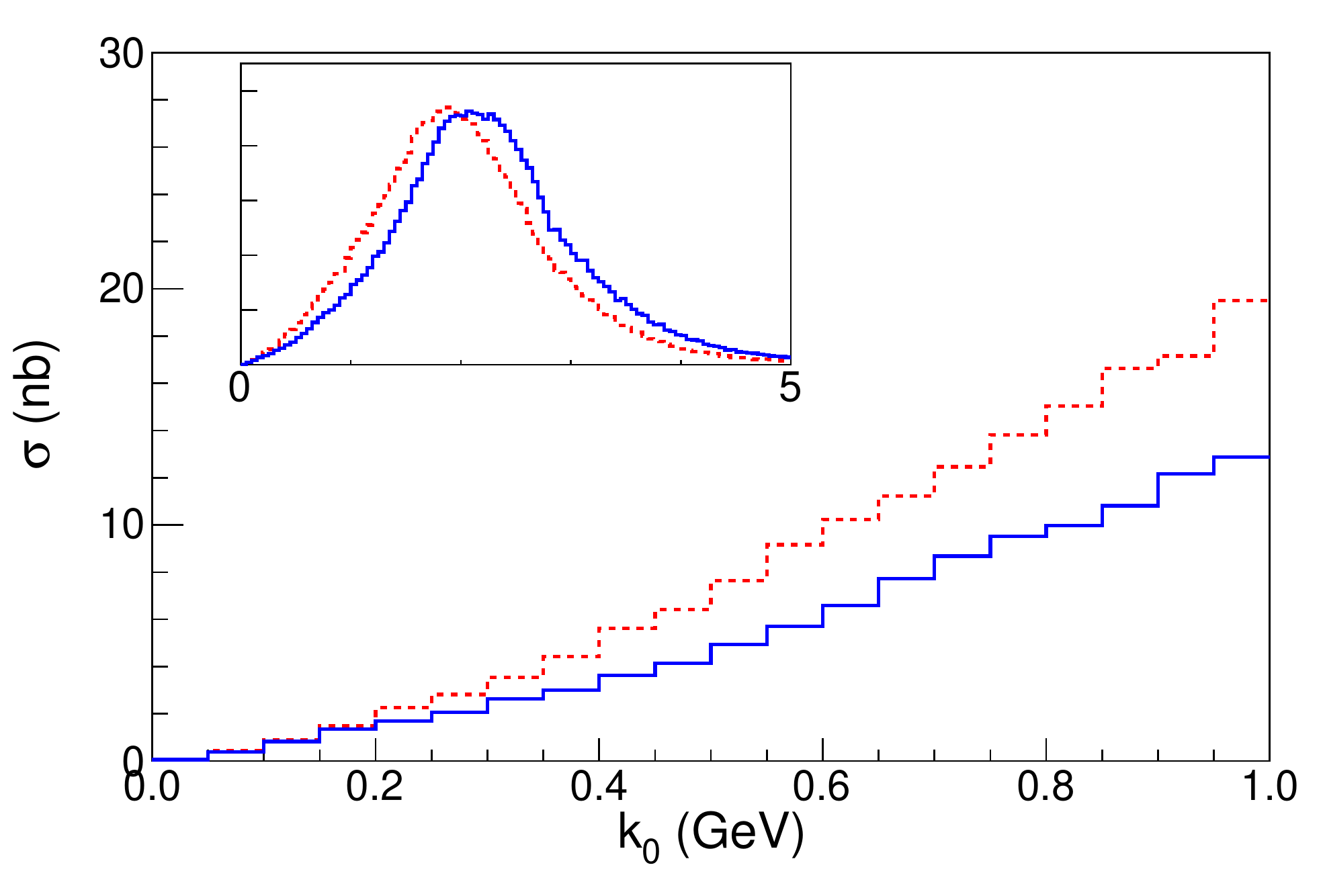}
 \caption{Integrated cross sections of $D^0 \bar D^{*0} + \text{h.c.}$ pairs at CDF, without (blue, solid) and with (red, dashed) the interaction with pions, as provided by HERWIG. The pion is selected as in~\cite{pioni}, with the additional constraint that $\Delta R_{DD^*} < 0.7$ before the interaction. In the inset the same plot on a wider range of $k_0$ values. The bin width here is larger by a factor of five than the one used in~\cite{bigna1}: the two plots are perfectly consistent. We use here a $K$ factor of 1.65 as in Table~I.  On the other hand the $K$ factor used in Fig.~2(a) of~\cite{arto} is apparently $K\approx 7$.}
\label{risultati}
\end{figure}

In Fig.~\ref{risultati} we show that allowing a single elastic scattering with a pion comoving with a $D$ or $D^*$ in the final state has the statistical effect of increasing the number of would-be molecules found in low energy bins, but not such to populate the $k_0\lesssim 50$~MeV region at the level required to fill the gap with the observed cross section -- which remains larger by two orders of magnitude. 

As shown above, the rescattering pion here considered is not spoiling the Monte Carlo agreement with the CDF $d\sigma/d\Delta\phi$ distribution  nor with $p_T$ distributions for open charm production we could compare to. 

The feed down from higher bins can be understood considering that the large majority of $D^0\bar D^{*0}$ pairs are found at high $k_0$ values (the maximum 
of the distribution is at about 2~GeV). Even if a small part of them is kinematically modified by the elastic scattering with a comoving pion in the direction of decreasing the com $k_0$, there would be a considerable shift of the distribution at lower $p_T$, but not such to populate the $k_0\in[0,50]$~MeV region. In order to prove this, we choose the interacting pion with the same recipe as in~\cite{pioni}: we randomly choose if a pion will interact with the $D$ or the $D^*$, then we select the 10 pions which are closest to the $DD^*$ plane, and finally we pick the pion the most parallel to the non-interacting meson. Moreover, we need to prevent that a $D$ and a $D^*$ of different jets (and probably far in the coordinate space) would be put closer by the scattering with a hard pion. For the interaction to happen, we thus require $\Delta R_{DD^*} = \sqrt{\left(\Delta y_{DD^*}\right)^2 + \left(\Delta \phi_{DD^*}\right)^2} < 0.7$; this additional cut does not modify the curve up to $2$~GeV. The selection 
procedure ensures that also the pion would be in the same jet.  

The open charm mesons might interact with pions more than once before a molecule is formed.
Roughly speaking the $\pi D^0\to \pi D^0$ scattering is proportional to $g_{{\pi D D^*}}^4$ whereas the $D^*\to D\pi$ decay is `slower' by  $g_{{\pi D D^*}}^2$.
We might say that $\tau_{\text{scatt}}\sim1/(\rho v \sigma)\approx M^2/(g^4 (200)^3 \text{MeV}^3)\approx 1/(10^4 g^2)\;  \text{MeV}^{-1}$ where we used the $\pi D$ reduced mass for $M$. On the other hand $\tau_{\text{dec}}\sim 24\pi M_D^2/(g^2 
|\bm p^*|^{3})$ where $|\bm p^*|$ is the decay momentum. Thus $\tau_{\text{dec}}\sim 10^4/g^2$~MeV$^{-1}$. 

 The hadronization time of a  $c$ or $q$ in the lab frame can be estimated to be of order~$t_{\text{had}}=({\cal E}/m_{c,q}) \; R$ where $R\approx 1$~fm and the mass is meant to be the constituent one. From our simulations we estimate that,  at the formation time of pions, the $D\bar D^{*}$ pair and  pions will be distributed on  {\it spherical segments} -- around $D$ and $D^*$ -- of an expanding sphere.  Using our simulations we estimate $2\div 3$ pions per~$\pi r_0^2$~fm$^2$ around $D$ or $D^*$, $r_0$ being the characteristic length scale of strong interactions $r_0\approx 3$~fm. 
\begin{figure}[h]
\centering
  \includegraphics[width=.48\textwidth]{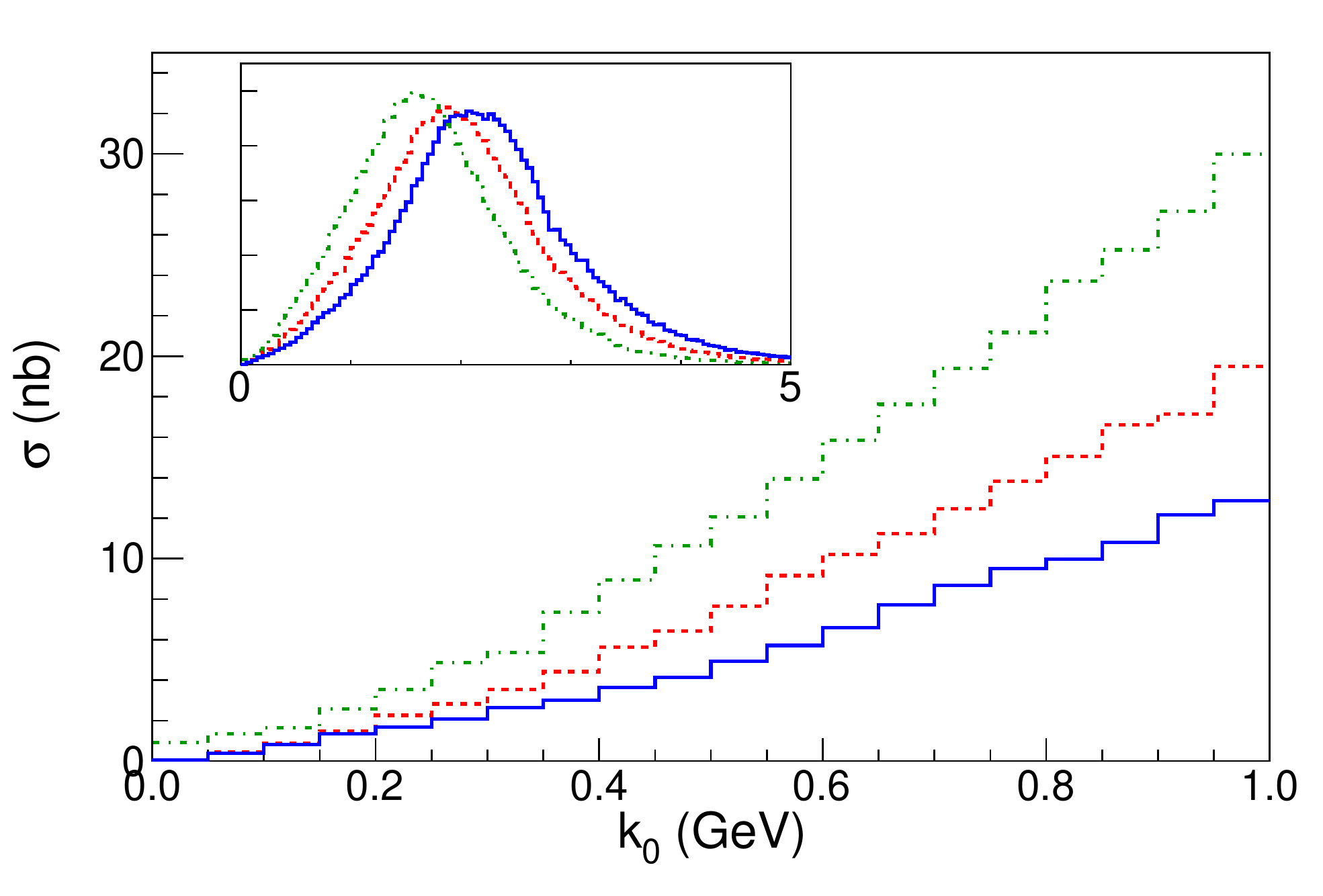}
\caption{Integrated cross sections of $D^0 \bar D^{*0} + \text{h.c.}$, without (blue, solid), with one (red, dashed) and three (green, dot-dashed) interacions with pions, with the method of~\cite{pioni}. In the insect the same plot on a wider range of $k_0$ values. }
\label{markov3}
\end{figure}
The results obtained with the method introduced in Ref.~\cite{pioni} applied to full QCD samples is shown in Table~\ref{tablemarkov} and Fig.~\ref{markov3}.

At any rate, we observe that elastic scatterings with comoving pions has a role  
and might be considered in all those attempts to evaluate cross sections of candidate molecular states. 
More specifically, including the interaction with a comoving pion, we get an enhancement of countings  
in the $k_0\lesssim 450$~MeV region such that the estimated cross section would be too large with respect 
to the observed one, see Table~\ref{tablemarkov}.  Also this might be relevant to the 
studies in~\cite{altri} and those to come.

\begin{table}[h]
\begin{tabular}{l|c|c|c}
$k_0^\text{max}$ & 50 MeV & 300 MeV & 450 MeV \\ \hline
$\sigma(0\pi)$ & 0.06 nb & 6 nb & 16 nb \\
$\sigma(1\pi)$ & 0.06 nb & 8 nb & 22 nb \\
$\sigma(3\pi)$ & 0.9 nb & 15 nb & 37 nb \\ \hline
\end{tabular}
\caption{Effect of multiple scattering in $X(3872)$ production cross section (see Fig.~\ref{markov3}). $k_0^{\rm max}$ indicates the integration range $[0,k_0^{\rm max}]$. To confront with the results obtained in~\cite{arto} we should multiply the cross section values by $7/1.65$, which is the ratio of $K$ factors used in the two papers (see caption in Fig.~4). For example using a $k_0^{\rm max}=360$~MeV, the $X$ prompt production cross section obtained in~\cite{arto} is $\sigma\approx 30$~nb (the experimental value at CDF), whereas, including the elastic  scattering with one pion, they would get $\sigma\approx 12.4*7/1.65=52$~nb where $\sigma(1\pi)=12.4$~nb for $k_0^{\rm max}=360$~MeV.}
\label{tablemarkov}
\end{table}
 
\section{Antideuteron data at ALICE}
We use ALICE preliminary data on antideuteron production at LHC: $p p$ collisions, $\sqrt{s}=7$~TeV, in the interval $0.4\text{ GeV} < p_T < 1.4\text{ GeV}$~\cite{alice} (see also E735 data about deuteron,~\cite{e735}). We generate samples of $2\to2$ full QCD events with HERWIG and PYTHIA. We checked that PYTHIA fails to reproduce the shape of the distribution in this low-$p_T$ region, so we use HERWIG for our analysis. We generated different MC samples ($1\times 10^8$ events each) with $\left|y^\text{max}(\text{part})\right| < 10$, and with different $p_T^\text{min}(\text{part})$ cuts, as in Table~\ref{tab:deuterio1}.
This test is needed to check the robustness of MC simulations when getting  close to $\Lambda_{\rm QCD}$. Our antideuteron candidates are the $\bar p \bar n$ pairs with $k_0 < 80~\text{MeV}$ (as suggested by the coalescence model in~\cite{strumia,arto}, and as predicted by a simple potential well model), and $\left|\eta\right|<0.9$. We rescale the curves in order to fit the data with $0.9\text{ GeV} < p_T < 1.4\text{ GeV}$. The Monte Carlo total cross section is very sensitive to the partonic $p_T$ cut, and the values obtained are much higher than the experimental value $\sigma\sim 90~\text{mb}$. However, we see that the number of 
antideuterons candidates has only a slight dependence on the cuts~(Table~\ref{tab:deuterio1}), and the shape of the distribution does not depend on them at all. 
\begin{table}[h]
\begin{tabular}{c|c|c}
  $p_{T}^\text{min}(\text{part})$ (GeV) & Total $\sigma$ (mb) & \# antideuteron candidates \\ \hline
 $2.0$ & $335$ & $47$k \\
 $1.6$ & $726$ & $50$k  \\
 $1.0$ & $1223$ & $55$k \\ \hline
\end{tabular}
\caption{Details of MC simulations as a function of $p_{T}^\text{min}(\text{part})$. We generate $10^8$ HERWIG full QCD events for each setup}
\label{tab:deuterio1}
\end{table}

Since different partonic cuts do not affect the distribution we get, we choose to analyze data with $p_{T}(\text{part}) > 2.0\text{ GeV}$, to stay in a safer region. We consider now different values of $k_0^\text{max}$: $80$~MeV (as suggested in~\cite{strumia,pioni}), $110$~MeV (as suggested in~\cite{arto}), and $300$~MeV (as suggested in~\cite{arto} for the $X(3872)$). We report the results in Table~\ref{tab:deuterio2}. We see that the number of deuteron candidates scales with $\left(k_0^\text{max}\right)^3$ as expected. However, the shape of the $p_T$ distributions is totally uncorrelated with $k_0^\text{max}$, so that the arbitrariness of $k_0^\text{max}$ is to some extent 
reabsorbed into the normalization factor. Hence, we can choose large values for $k_0$ ($300-450$~MeV) to improve statistics without affecting the $p_T$ distribution.

\begin{table}[h]
\begin{tabular}{c|c}
  $k_\text{0}^\text{max}$ (MeV) & \# antideuteron candidates \\ \hline
 $80$ & $454$k\\
 $110$ & $1.2$M \\
 $300$ & $22.5$M \\ \hline
\end{tabular}
\caption{Details of MC simulations as a function of $k_0^\text{max}$. We generate $10^9$ full QCD events with HERWIG.}
\label{tab:deuterio2}
\end{table}

\begin{figure}[h]
\centering
 \includegraphics[width=.48\textwidth]{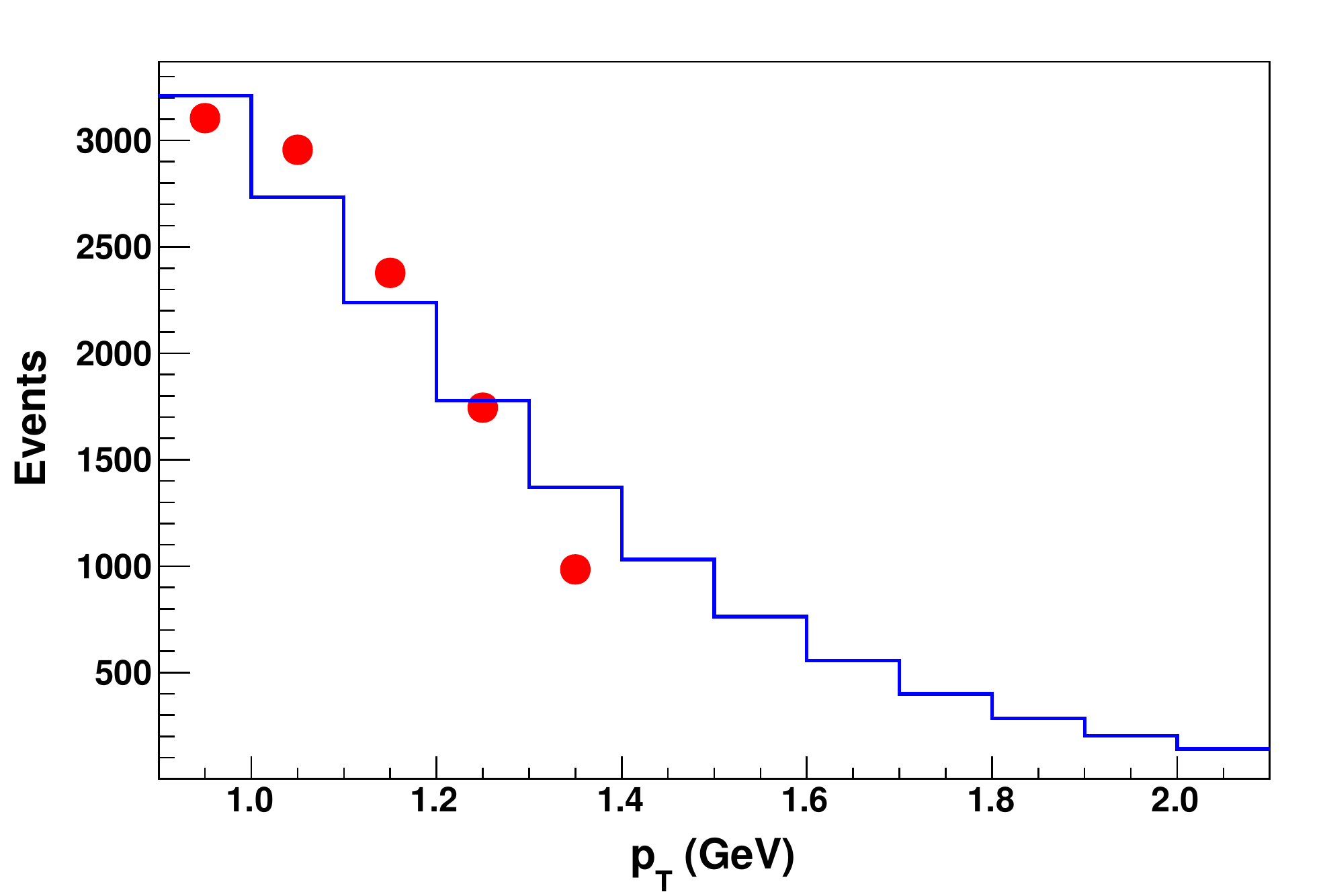}
 \includegraphics[width=.48\textwidth]{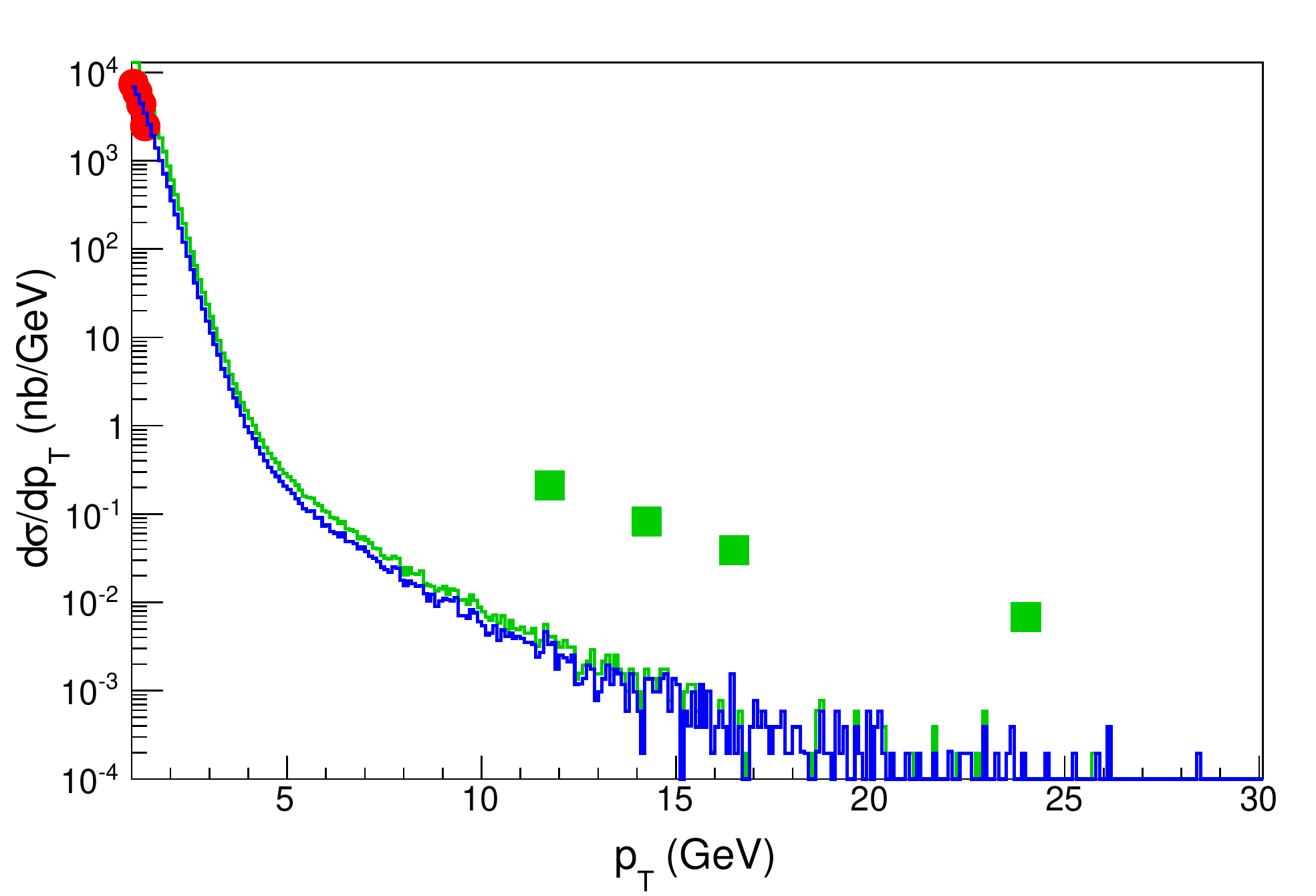}
 \caption{Antideuteron events produced in $p p$ according to $10^9$ HERWIG events. We confront with ALICE deuteron production data (red circles)~\cite{alice}, and with CMS $X(3872)$ data (green squares)~\cite{cms}. The blue solid line is the MC prediction in the $\left|\eta\right|<0.9$ region, as in  ALICE data, which we use for the normalization.  The green line (a bit higher in the right panel) corresponds to the $\left|y\right|<1.2$ region, as in CMS data, and is normalized accordingly.  }
\label{antideuterio}
\end{figure}

With this tuning, we can study the MC distribution in the high $p_T$ region. 
We see that we are about three orders of magnitudes below the experimental cross section of the $X(3872)$ measured by CMS~\cite{cms}. 
In these respects we are assuming that the $X(3872)$ is a kind of mesonic deuterium and that spin-interactions play little role. Anyway there is a qualitative trend, as can be appreciated in Fig.~\ref{antideuterio}, that is quite suggestive. As a caveat, consider that the extrapolation in $p_T$ we are doing here is very broad, that MC predictions at such small $p_T$ values are affected by large uncertainties, and that ALICE data are still preliminary and not efficiency-corrected. Anyway, more data from ALICE in the intermediate region, say up to 3 to 5 GeV, could improve the reliability of this extrapolation.

\begin{figure}[h]
\centering
 \includegraphics[width=.48\textwidth]{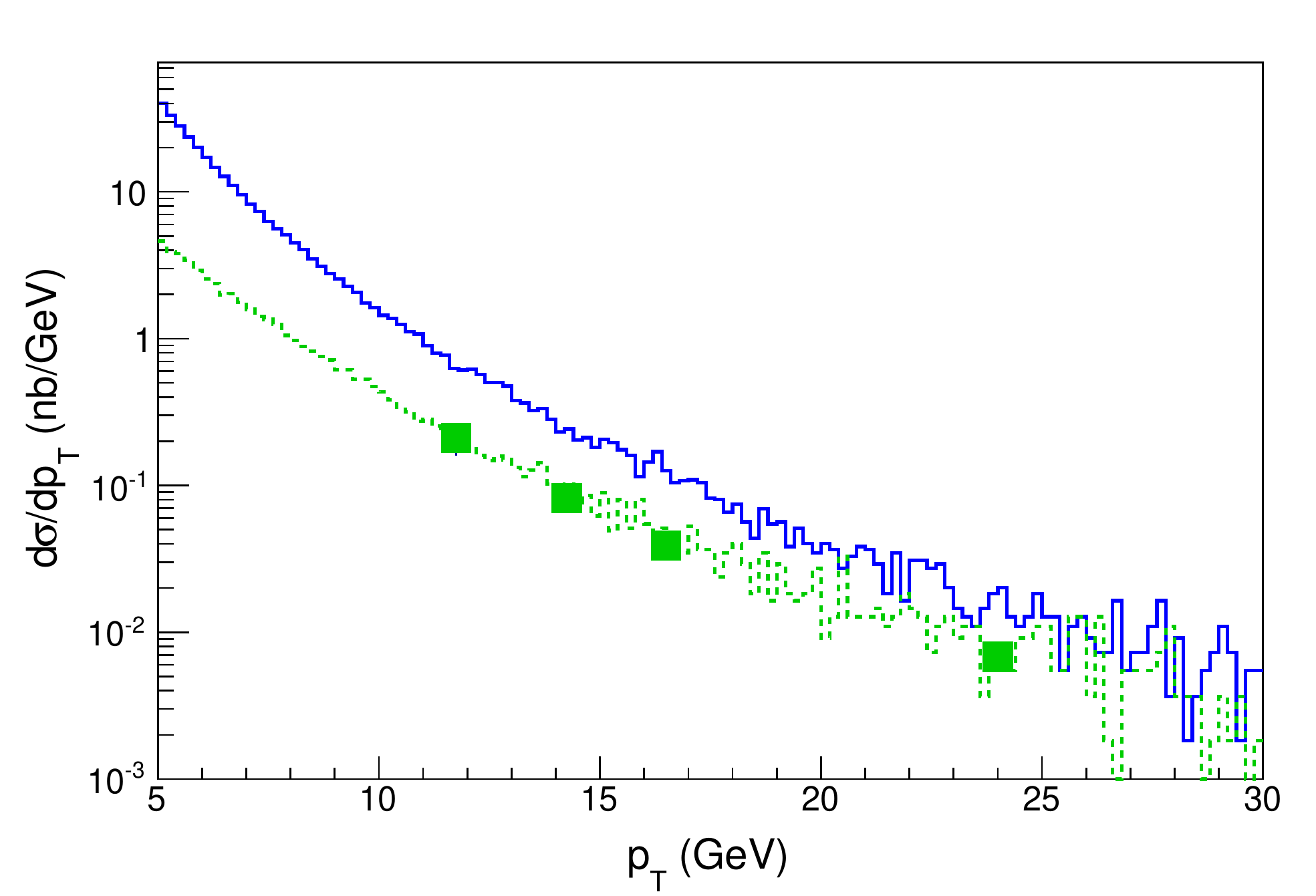}
 \includegraphics[width=.48\textwidth]{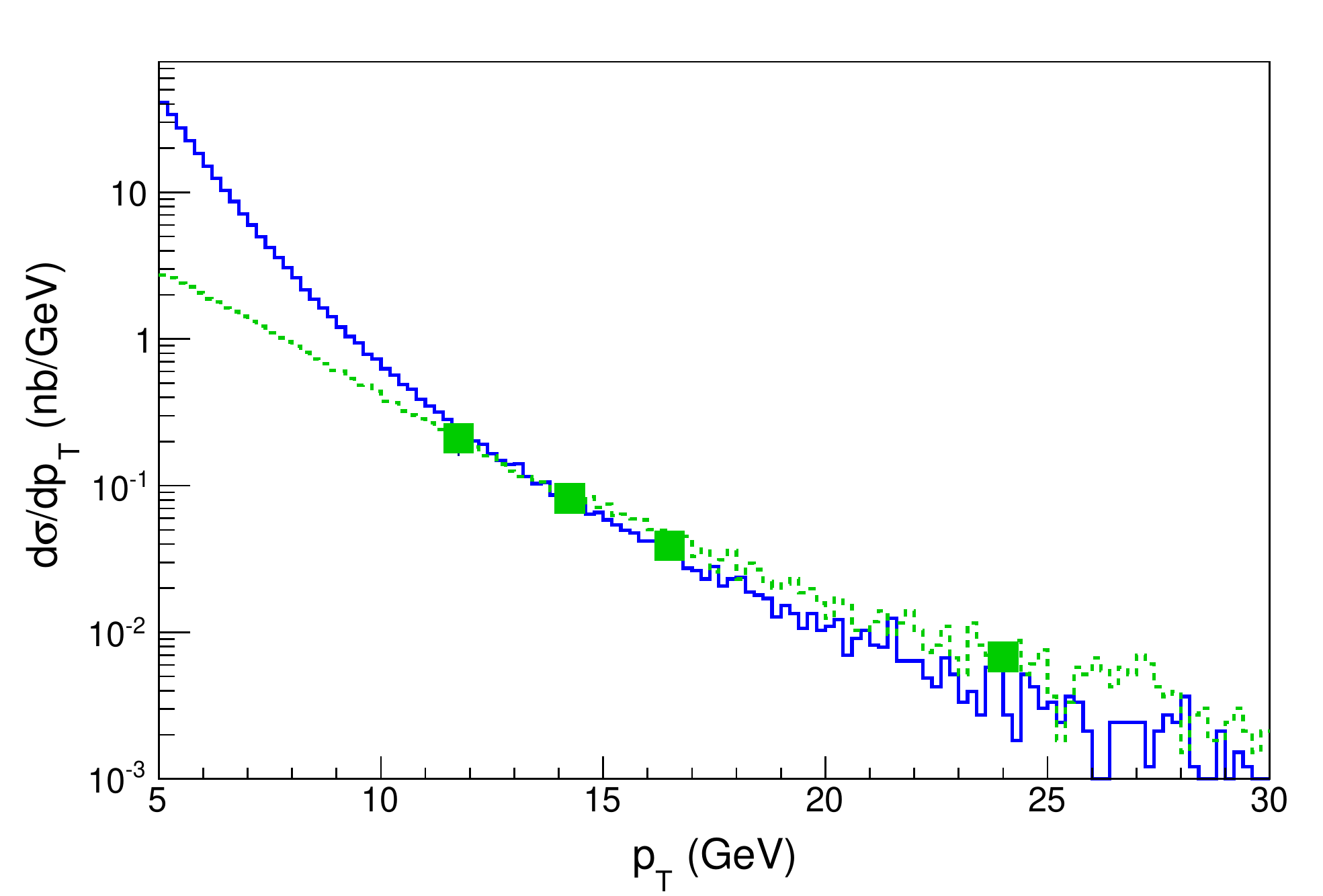}
\caption{Antideuteron (blue, solid) and $X(3872)$ (green, dashed) candidates. We generate $20\times 10^9$ events with HERWIG (left panel) and PYTHIA (right panel), with partonic cuts $p_T^\text{part} > 3.5$~GeV and $\left|y^\text{part}\right| < 6$. We select pairs with $k_0 < 450$~MeV, and with the experimental cuts $\left|\eta\right|<0.9$ for antideuteron~\cite{alice} and $\left|y\right|<1.2$ for the $X(3872)$~\cite{cms}. The curves are normalized according to the experimental points of the $X(3872)$ (green squares,~\cite{cms}). We see that with this normalization we predict the antideuteron curve to be larger by two orders of magnitude with respect to the blue curve in Fig.~\ref{antideuterio} normalized according to ALICE antideuteron data at low $p_T$~\cite{alice}. }
\label{gambero}
\end{figure}

To further investigate this matter we relate 
the predicted production cross sections 
for the $X(3872)$ and antideuteron through their ratio in the (perturbative) 
$p_T$ range where the $X(3872)$ is observed at  CMS. 
The Monte Carlo prediction is shown in Fig.~\ref{gambero}, 
where both the HERWIG and PYTHIA results are shown (left and right panel of Fig.~\ref{gambero}, 
respectively). The distribution of $X(3872)$ 
is normalized according to data. No data are available for antideuteron in this range, 
so no direct comparison is possible. 
We find that, at $p_T\sim 5$~GeV, the antideuteron curve in Fig.~\ref{gambero} is two orders 
of magnitude larger than the one in Fig.~\ref{antideuterio}. 
Furthermore, even without trusting the large extrapolation of Fig.~\ref{antideuterio}, 
the $X(3872)$ production cross section in the large $p_T$ region appears 
to be limited by the antideuteron production cross section, Of course also this prediction 
could be affected by large uncertainties in the Monte Carlo modelling of antiproton and 
antineutron production w.r.t. charmed meson pairs (the difference between HERWIG and PYHTIA 
curves gives a first estimate of these uncertainties). For this reason we remark 
the importance of a measurement of antideuteron production in the $p_T$ range of 
the $X(3872)$, in order to fix the ambiguities of the Monte Carlo predictions. 
Should be confirmed the picture of a much lower antideuteron production 
w.r.t $X(3872)$, it would challenge the hypothesis for this two states to share the same nature. 

The qualitative conclusion we might draw by the present analysis is that there is indeed a production rate of antideuteron, a `loosely' bound baryon molecule, but only within a low transverse momentum region: if we extend the search at higher transverse momenta, where the single components are allowed to recoil harder from each other, we will find much less antideuteron. Indeed, to make an example,  if $\bm p_1$ and $\bm p_2$ are the 3-momenta of $\bar p$ and $\bar n$ and we assume for simplicity that   $|\bm p_1|\approx|\bm p_2|$, $\varphi$ being the angle between them, the relative momentum in the lab is orthogonal to the boost $|\bm p_1+\bm p_2|/({\cal E}_1+{\cal E}_2)$  and we can estimate $k_0\approx p_T \tan(\varphi/2)$, where $p_T=|\bm p_1+\bm p_2|\sin\theta$, assuming the angle $\theta$ to the beamline being non-negligible (cuts we use in rapidity give $\theta \approx 45^\circ$).   For generic $\varphi$ angles, higher $p_T$ values drive higher $k_0$ values.

\section{The diquark-antidiquark description and Hadronization}
On the basis of the discussion reported in the previous sections, we are led to conclude that the hadronization of multiquark hadrons in prompt collisions at LHC must proceed through the formation of compact quark clusters,
with color neutralized in all possible ways. A $|\psi\rangle=|Q\bar Qq\bar q\rangle$ state is therefore a superposition of the alternative color configurations (leaving aside, for the sake of simplicity, repulsive channels in the one-gluon-exchange model)
\begin{equation}
|\psi\rangle =\alpha |[Qq]_{\bar{\bm 3}_c}[\bar Q\bar q]_{\bm 3_c}\rangle+\beta |(Q\bar Q)_{\bm 1_c}(q \bar q)_{\bm 1_c}\rangle + \gamma |(Q\bar q)_{\bm 1_c}(\bar Q q)_{\bm 1_c}\rangle 
\label{superp0}
\end{equation}

The two-meson states will tend to fly apart, as strong Van der Waals-like forces between their meson components are not sufficient to produce bound states like $J/\psi \,\rho$ or $D\bar D^*$, depending on the spin and orbital quantum numbers of the original four-quark system.  In this sense such  states are in a `open channel'. We might describe them as levels at the onset of the continuum spectrum of some very shallow (strong Van der Waals) potential. 

The diquark-antidiquark state is instead kept together by color interactions, the unknown being the effectiveness of the color force at producing diquarks. The one-gluon exchange model would qualitatively  suggest that the attraction in the diquark channel is just a factor of two less intense than in the quark-antiquark channel. In this sense we define a diquark-antidiquark state to belong to a `closed' channel. Thus we may write
\begin{equation}
|\psi\rangle =\alpha |[Qq]_{\bar{\bm 3}_c}[\bar Q\bar q]_{\bm 3_c}\rangle_{{\mathcal C}}+\beta |(Q\bar Q)_{\bm 1_c}(q \bar q)_{\bm 1_c}\rangle_{{\mathcal O}} + \gamma |(Q\bar q)_{\bm 1_c}(\bar Q q)_{\bm 1_c}\rangle_{{\mathcal O}} 
\label{superp}
\end{equation}
where ket subscripts ${\mathcal C},{\mathcal O}$ indicate `closed' and `open' channels respectively.
The projection operators ${\mathcal C}$ and ${\mathcal O}$ onto the respective Hilbert spaces are such ${\mathcal C}{\mathcal O}=0$, thus the states in~(\ref{superp}) are to be considered as othogonal.   The relative size of $\alpha,\beta,\gamma$ coefficients is unknown. 

We might formulate different
hypotheses: $i)$ $\alpha,\beta,\gamma$ are all of the same order. In this case we should be observing the entire spectrum of diquark-antidiquark states.  

Models to compute this spectrum stem from assumptions on spin-spin interactions between quarks. In its first version~\cite{xmai}, the Hamiltonian of the diquark-antidiquark model was supposed to contain both spin-spin interaction between quarks within each diquark and quarks outside the single diquark shells. The resulting spectrum predicts a rich structure of states with some evident mismatches with the most recent experimental findings. 

A `type-II' version of the diquark-antidiquark model, with  spin couplings suppressed between different diquarks, allows  a remarkable description of the $J^{PG}=1^{++}$ sector of charged tetraquarks as the $Z(4430)$, $ Z_c(4020)$, $Z_c(3900)$ and a very suggestive picture of the entire $J^{PC}=1^{--}$ sector~\cite{wiprog}.  Some typical problems of the diquark-antidiquark model persist in the type-II  model. For example the $X(3872)$ should have charged partners and an hyperfine splitting between two neutral levels, to account for isospin violations. 

To solve this kind of difficulties we might formulate a different hypothesis for the hierarchy among $\alpha,\beta,\gamma$ coefficients. 
We might indeed assume that,  $ii)$
\begin{equation}
|\beta|^2,|\gamma|^2 \gg |\alpha|^2
\end{equation}

Such an assumption means that, in general, diquark-antidiquark states are  less likely to be formed in hadronization but a resonance could emerge as a result of the coupling between open and closed channels. This hypothesis introduces a selection rule in the diquark-antidiquark spectrum: 
especially those levels which are close enough to open channel levels (resonance conditions) are observed as physical resonances.
 
More specifically, the diquark-antidiquark {\it closed} channel provides an effective attraction in the open channel which might lead to produce a resonance. This phenomenon is effective if the energy level $E_n$, corresponding to the closed channel state $|[Qq]_{\bar{\bm 3}_c}[\bar Q\bar q]_{\bm 3_c},n\rangle_{{\mathcal C}}$, happens to be very close to one, or both, as in the $X$-particle case, of the open channel thresholds, located at $E_{\mathcal O}=m_{J/\psi}+m_\rho$ or $E_{\mathcal O}=m_{D^0}+m_{\bar D^{*0}}$.   

Strong force provides the discrete spectrum for diquark-antidiquark states, however 
those levels correspond to physically realized states once the closed channel is hybridized with the open one, {\it i.e.} the difference in energy, or detuning parameter $\nu$, is small enough.  When this energy matching condition between the total energy in the open channel and the energy level in the closed channel takes place, the two hadrons in one open channel can undergo an elastic scattering, altered by the presence of the near closed channel level. The two hadrons in an open channel can scatter to the intermediate state in the closed channel, which subsequently decays to give two particles in one of the open channels. 
 
 The contribution to the scattering length due to this phenomenon is of the form 
 \begin{equation}
 a\sim |C|\sum_n \frac{_{{\cal C}}\langle [Qq]_{\bar{\bm 3}_c}[\bar Q\bar q]_{\bm 3_c}, n | 
 H_{{\cal C}{\cal O}}|(Q\bar q)_{\bm 1_c}(\bar Q q)_{\bm 1_c}\rangle_{{\mathcal O}} }{E_{{\cal O}}-E_{n}}
 \label{fesh}
 \end{equation}
This sum is dominated by the term which minimizes the denominator $E_{{\cal O}}-E_{n} \equiv -\nu$, {\it i.e.}  the one with the smallest detuning. The width of the resulting resonance is naturally proportional to the detuning $\Gamma\sim \sqrt{\nu}$ for phase space arguments.

Since the $X(3872)$ is the narrowest among all {\xyz} mesons, it must have  $\nu\approx 0$, which means the highest possible hybridization between channels given the (unknown) inter-channel interaction Hamiltonian $H_{{\cal C}{\cal O}}$. The input value to fix the discrete closed channel diquark-antidiquark spectra is  the mass of the $X(3872)$: we require that the lowest $1^{++}$ state has the mass of the $X$ and this fixes the diquark mass and the spectra as in~\cite{xmai} or~\cite{wiprog}. 
 
The $D^+D^{*-}$ open threshold is found  to be at a mass {\it above} the $X$ diquark-antidiquark level, by about 8~MeV. Coupling between channels gives rise to a repulsive interaction if the energy of the scattering particles is larger than that of the bound state (and an attractive one if it is smaller).  For this reason we might conclude that the neutral particle has no $d\bar d$ content in its wavefunction explaining the well known isospin breaking pattern in  $X$ decays.  

The diquark-antidiquark $X^{+}$ levels, might also fall below $D^+\bar D^{*0}$ and $\bar D^0 D^{*+}$ thresholds by about $3\div 5$~MeV, which could be enough for inhibiting the resonance phenomenon described. This could be the reason why the $X^{+}$ particles, although present in the diquark-antidiquark spectrum, are more difficult to be produced.

The $J/\psi\;\rho^0$ open channel level is also perfectly matching the closed channel one for the $X(3872)$. However because of the large $\rho$ width, the modification in the scattering lenghth~(\ref{fesh}) is much less effective if compared to the open charm threshold: the sum in~(\ref{fesh}) has to be smeared with an integral convoluting the $\rho$ Breit-Wigner.  Therefore we would expect that the $X^{+}$ particles are less likely to be formed or they could simply be too broad to be observed.

The mechanism here described is known in nuclear and atomic physics as the Feshbach resonance formation~\cite{feshb}. 

Recently two {\it charged} resonances have been confirmed to a high level of precision. The $Z(4430)$~\cite{z4430} and the $Z_c(3900)$~\cite{bes3}. These are genuine tetraquark states. 

We have to remind here that charged states were qualitatively predicted only by the tetraquark model in~\cite{xmai}. In particular, when the first hint of a $Z(4430)$ charged tetraquark was provided by the Belle Collaboration~\cite{z4430belle} in the $\psi(2S)\,\pi^+$ channel, back in 2007, it was observed that another state  at 3880 MeV ({\it i.e.} lighter by the $\psi(2S)-\psi(1S)$ mass difference) was expected in the tetraquark model~\cite{tetraz} with the same quantum numbers (the former being the radial excitation of the latter). In particular the lower state was expected to decay into 
$J/\psi \,\pi^{+}$ or $\eta_c \, \rho^{+}$ -- this was a  prediction of the diquark-antidiquark model~\cite{xmai,tetraz}. Just last year a charged $Z_c(3900)$ with $J^{PG}=1^{++}$ decaying into $J\psi\,\pi^+$ was discovered by BES~III and Belle~\cite{bes3}; see also~\cite{xmai2}. 

The tetraquark model in its first diquark-antidiquark version~\cite{xmai,xmai2} predicts one more $J^{PG}=1^{++}$ level, at a mass of $3755$~MeV (these mass values are locked to the input mass value of the $X(3872)$).  We might predict that no resonance will be found at this level because there are no open channels nearby to make the Feshbach mechanism possible. The $Z(4430)$ is instead made possible by the presence of the $\eta_c(2P)\rho$ open channel. The expected width, driven by the $\rho$, is expected to be as large as $150$~MeV, to be compared with the $\sim 170$~MeV observed. 

The tetraquark model in its `type-II' version has no $3755$~MeV, but a level perfectly compatible with the observed $Z_c^\prime (4020)$ by the BES~III Collaboration~\cite{4020}, which is also compatible with a Feshbach generated state. 

A $Z(4430)^0$ isosinglet resonance could be due to the vicinity of the $\eta_c(2P)\;\omega$ open channel, with a narrower width of about $70$~MeV. 

The $Z_c(3900)$ and $Z(4430)$ should also be promptly produced at LHC. We expect a smaller production cross section for the $Z(4430)$ if compared to the $Z_c(3900)$, which happens to be  closer to the related threshold. Similarly for the $J^{PC}=1^{--}$ states, aka $Y$  states, whose prompt production cross sections are also expected to be small because not amplified by the Feshbach resonance  mechanism described above. This expectation is supported by the model-independent estimate provided by Ali and Wang~\cite{ali2}, from which we can infer that
\begin{equation}
\frac{\sigma(pp\to Y(4260))}{\sigma(pp\to X(3872))}\lesssim 10^{-2}\end{equation}
confirming our suggestion of a small $\alpha$ parameter in Eq.~(\ref{superp0}). The inequality comes from the fact that ${\cal B}(X(3872)\to J/\psi\,\pi^+\pi^-)\approx5\%$~\cite{noirev}, whereas the same branching fraction for the $Y(4260)$ is expected to be equal or larger (say $\gtrsim 1\%$ as in~\cite{olsen}). 

For the moment we do not extend our  analysis to tetraquark states with hidden strangeness~\cite{ssbar} like the $Y(4140)$~\cite{vistoda}.

Recent experimental findings are clearly spelling in favor of tetraquark particles and the diquark-antidiquark model apparently has many features matching very well the present phenomenology. The Feshbach mechanism here sketched might be a viable  way for  implementing selection rules in the tetraquark Hamiltonian. On the basis of what discussed here we understand that the measurement of prompt production cross sections of all {\xyz} resonances will be of crucial importance to help theoretical interpretations. 

\section{Conclusions}
Monte Carlo simulations strongly disfavor the production of loosely bound open charm molecules in prompt collisions at hadron colliders when reasonably small relative momenta in would-be-molecules are allowed. This holds true even if we allow elastic scatterings with final state pions comoving with a $D^0$ or $D^{*0}$ component of the would-be $X(3872)$ molecule. 
 Final state rescatterings with pions do not spoil the Monte Carlo tuning on data or $p_T$ distributions of open charm meson production. 
 
 On the other hand, elastic final state interactions with one or more (up to three) pions sensibly change the expected values for molecule prompt production cross sections
 when higher relative momenta between molecular components are assumed, say $k_0\in[300, 500]$~MeV.
This could be relevant to those studies attempting to estimate {\xyz} prompt production cross sections at LHC using the loosely bound molecular hypothesis. We also use different 
Monte Carlo techniques to produce the final state, to review all the features and possible loopholes of this method.
 
In the same context a  preliminary study of some ALICE data on antideuteron production at the LHC are studied, using the Monte Carlo method to extrapolate in $p_T$ up to the transverse momentum region where the $X$ is better observed by CMS. This is done with the purpose of a qualitative comparison between the $X$ and deuteron, often addressed as its baryonic analogous loosely bound state.  More data in the intermediate $p_T$ region would be of help in making this extrapolation more quantitative. Our results are consistent with a very poor expected production of antideuteron at high $p_T$, orders of magnitude lower than the abundant $X$ production.

We suggest an alternative way of looking at the formation of {\xyz} particles whose production might be amplified by a Feshbach resonance phenomenon in the open channels of charmed meson pairs induced by the closed channel of diquark-antidiquark levels. The relevance of the diquark-antidiquark model to the {\xyz} phenomenology is strongly suggested by recent experimental findings.
This approach introduces some  selection rules in the diquark-antidiquark model reducing the number of expected states. The features of the $X$ seem to fit reasonably well  in our  scheme and the newly discovered charged resonances $Z(4430)$ and $Z_c(3900)$ are also well understood. We provide predictions for their expected prompt production cross sections and for the width of the neutral $Z(4430)^0$, yet to be experimentally confirmed. As a last comment, the existing model-independent estimates of the $Y(4260)$ 
production cross section at the LHC support the proposed scenario for the formation of 
{\xyz} particles. 
 
 {\bf \emph{Acknowledgements}} We whish to thank M. Bochicchio, C. Bonati, M. Caselle,  M. D'Elia, A. Esposito, L. Maiani, R. Mussa, R. Nania, M. Papinutto, V. Riquer,  N. Tantalo for many stimulating discussions and suggestions.

\end{document}